\pdfoutput=1
\documentclass[conference]{IEEEtran}

\usepackage{etex}
\usepackage{amssymb}
\usepackage{times}
\usepackage{bm}
\usepackage{subcaption}
\usepackage{amsmath}
\usepackage{graphicx}
\usepackage{amssymb}
\usepackage{amstext}
\usepackage{latexsym}
\usepackage{pifont}
\usepackage{color,soul}
\usepackage{ifthen}
\usepackage{multirow}
\usepackage[linesnumbered,ruled]{algorithm2e}
\usepackage{verbatim}
\usepackage{array,tabularx}
\usepackage{accents}
 \usepackage{cite}
 \usepackage{arydshln}
\usepackage{hyperref}
\usepackage{url,amssymb,epsfig,times,mathrsfs,algorithmic,latexsym,xspace,fancyhdr,wrapfig, xcolor,multirow,amsthm}
\usepackage{caption}
\usepackage{soul}
\usepackage[inline]{enumitem}
\usepackage{arydshln}
\usepackage{enumitem}
\usepackage[makeroom]{cancel}
\usepackage{mathtools}
\usepackage{dblfloatfix}
\usepackage[utf8]{inputenc}

\hypersetup{nolinks=true}
\usepackage{tikz}
\usetikzlibrary{shapes.geometric}
\usetikzlibrary{shapes,snakes,patterns}
\usetikzlibrary{arrows,positioning,automata,calc}
\usetikzlibrary{plotmarks}
\tikzset{
    int/.style={
           rectangle,
           rounded corners,
           draw=black, thin, fill=black!20,
           minimum height=2em,
           inner sep=2pt,
           text centered,
           },
}

 \newcommand{\lev}{\text{lev}}

\newtheorem{theorem}{Theorem}

\newtheorem{claim}{Claim}
\newtheorem{proposition}{Proposition}
\newtheorem{corollary}{Corollary}
\newtheorem{remark}{Remark}

\newtheorem{construction}{Construction}

\IEEEoverridecommandlockouts

\pdfoutput=1

\begin{document}
\pdfoutput=1
\allowdisplaybreaks
\newlength\figureheight
\newlength\figurewidth

\title{Coding for Trace Reconstruction over Multiple Channels with Vanishing Deletion Probabilities}
\author{
\IEEEauthorblockN{Serge Kas Hanna }

\IEEEauthorblockA{ 
Institute for Communications Engineering, Technical University of Munich, Germany \\
Email: serge.k.hanna@tum.de} 
\thanks{This project has received funding from the European Research Council (ERC) under the European Union?s Horizon 2020 research and innovation programme (grant agreement No. 801434).}
}

\maketitle
\begin{abstract} 
Motivated by DNA-based storage applications, we study the problem of reconstructing a coded sequence from multiple traces. We consider the model where the traces are outputs of independent deletion channels, where each channel deletes each bit of the input codeword \(\mathbf{x} \in \{0,1\}^n\) independently with probability \(p\). We focus on the regime where the deletion probability \(p \to 0\) when \(n\to \infty\). Our main contribution is designing a novel code for trace reconstruction that allows reconstructing a coded sequence efficiently from a constant number of traces. We provide theoretical results on the performance of our code in addition to simulation results where we compare the performance of our code to other reconstruction techniques in terms of the edit distance error.
\end{abstract}

\section{Introduction}
DNA-based storage systems are celebrated for their ultrahigh storage densities that are of the order of \mbox{$10^{15}$-$10^{20}$} bytes per gram of DNA~\cite{DNA}. These systems also introduce various challenges including, but not limited to, data degradation due to DNA aging and errors introduced during DNA sequencing. The aforementioned challenges often lead to deletion, insertion, and substitution errors in the stored data~\cite{RO,DNAcoded}. For instance, DNA sequencing with nanopores results in multiple erroneous reads of the data. As a result, recovering the data in question can be cast in the setting of {\em trace reconstruction} which is a well-studied problem in the computer science community that was introduced in~\cite{Batu}. 

In trace reconstruction, the goal is to reconstruct an unknown sequence $\mathbf{x}$ given random traces of $\mathbf{x}$, where each trace is generated independently as the output of a random deletion channel. The deletion channel deletes each bit of $\mathbf{x}$ independently and with probability~$p$. The goal is to devise efficient algorithms that allow reconstructing $\mathbf{x}$ from a few traces. There have been lots of works in the literature that study trace reconstruction in the case where $\mathbf{x}$ is uncoded \cite{n1,n2,Batu,holenstein2008trace,mcgregor2014trace,de2017optimal,nazarov2017trace, hartung2018trace,peres2017average,holden2018subpolynomial}. 

Motivated by the application to DNA-based storage, we study the problem of trace reconstruction in the coded setting, where we are allowed to encode the sequence $\mathbf{x}$ before observing multiple traces of it.  Encoding data before storing it is a natural and widely-used strategy to ensure data reliability. However, employing this strategy in DNA-based storage systems requires novel techniques due to the uniqueness of such systems in terms of the methods used to read the data and also the types of errors that are experienced. Encoding data in DNA-based storage can be achieved by employing DNA synthesis strategies that enable encoding arbitrary digital information and storing them in DNA~\cite{church2012next,goldman2013towards,yazdi2017portable}. 

One way to enhance reliability in DNA-based storage is to use deletion correcting codes (e.g.~\cite{B16,GC,H19,Chen18,SimaIT,SimaSYS}), which would only require a single trace to reconstruct the stored data. However, using a deletion correcting code to correct $d$ deletions in the stored data would optimally require a redundancy (i.e., storage overhead) that is $\Theta(d \log (n/d))$~\cite{L66}; moreover, most of the existing deletion codes in the literature have complex encoding and decoding algorithms. Therefore, in our study on coded trace reconstruction, our goal is to exploit the presence of multiple traces which arises naturally in DNA-based storage applications (e.g., nanopore sequencing), to design codes that can be efficiently reconstructed from few traces, and have a redundancy that grows strictly slower than the optimal redundancy of deletion correcting codes.

The work by Abroshan {\em et al.} in~\cite{AbroshanTrace} studies coded trace reconstruction in the setting where each trace is affected by $d$ deletions, where $d$ is fixed with respect to~$n$ and the positions of the $d$ deletions are uniformly random. The high-level idea of the approach in~\cite{AbroshanTrace} is to concatenate a series of Varshamov-Tenengolts (VT)~\cite{VT65} codewords, which are single deletion correcting codes that have an asymptotically optimal redundancy in $n$; in addition to efficient linear time encoding and decoding algorithms. The number of concatenated VT codewords is chosen to be a constant that is strictly greater than $d$ so that the average number of deletions in each VT codeword is strictly less than $1$. The reconstruction algorithm then decodes each VT codeword by either finding a deletion-free copy of it among the traces or by decoding a single deletion using the VT decoder. Since the length of each codeword is a constant fraction of $n$, and since the redundancy of VT codes is logarithmic in the blocklength, then it is easy to see that the resulting redundancy of the code in~\cite{AbroshanTrace} is $\Theta(\log n)$. The works in~\cite{DNAcoded} and \cite{B20} study coding for trace reconstruction in the regime where the deletion probability $p=\epsilon$ is fixed, i.e., a linear number of bits are deleted on average. For this regime, the authors in~\cite{DNAcoded} introduced binary codes with rate $1-O(\frac{1}{\log n})$ that can be efficiently reconstructed from $\log^cn$ traces, with $c>1$. The authors in~\cite{B20} prove the existence of binary codes with rate $1-O(\frac{1}{\log n})$ that can be reconstructed from $\exp \left( (\log \log n)^{1/3} \right)$ traces. 

In this paper, we focus on coded trace reconstruction for binary sequences affected by deletions only, as a first step towards designing codes that are robust against the various types of errors that are experienced in DNA-based storage. We consider the random deletion model where each bit is deleted independently with probability $p$. We focus on the regime where $p=k/n^{\alpha}$, where $k>1$ and $\alpha \in (0.5,1]$ are constants. We design a novel code $\mathcal{C}\subset \{0,1\}^n$ for trace reconstruction and we present the following contributions.
\begin{itemize}[leftmargin=*]
\item For all $\alpha\in (0.5,1]$, the redundancy of our code is $o(n^{1-\alpha} \log n)$, i.e., grows strictly slower than the optimal redundancy of a deletion correcting code, which is $\Theta(n^{1-\alpha} \log n)$ for $d=\Theta(np)=\Theta(n^{1-\alpha})$. Furthermore, for the particular regime where the average number of deletions is fixed, i.e., $\alpha=1$ and $p=k/n$, we show that our code can efficiently reconstruct any codeword $\mathbf{x}\in \mathcal{C}$ with any redundancy that is $\omega(1)$. This regime is similar to the regime where the number of deletions per trace is fixed, which was studied in~\cite{AbroshanTrace}. For this regime, our code has a lower redundancy compared to the $\Theta(\log n)$ redundancy of the code introduced in~\cite{AbroshanTrace}.

\item We provide a theoretical guarantee on the performance of our code in the form of an upper bound on the probability of reconstruction error. This bound shows that for all $\alpha\in (0.5,1]$, and for any codeword $\mathbf{x}\in \mathcal{C}$, the probability of reconstruction error vanishes asymptotically in $n$ for a {\em constant} number of traces $t=O(1)$. 

\item We evaluate the numerical performance of our code in terms of the edit distance error, which was adopted as a performance metric in several recent related works~\cite{Diggavi, Eitan, Approximate}. We compare the performance of our code to the code in~\cite{AbroshanTrace} and to a simple coded version of the Bitwise Majority Alignment (BMA) reconstruction algorithm~\cite{Batu}. The simulation results show that our code has a low edit distance error for a small number of traces and outperforms both the code in~\cite{AbroshanTrace} and a simple coded version of the BMA algorithm. 

\item Any codeword can be reconstructed from traces in linear time $O(n)$.
\end{itemize}

The paper is organized as follows. We introduce the deletion model and some notations in~Section~\ref{sec:4:2}. In Section~\ref{sec:4:3}, we present our code construction and state our theoretical results. The simulation results are presented in Section~\ref{sec:simul}.
%
%

\section{Preliminaries}
\label{sec:4:2}

We consider the following deletion model. For a given input sequence $\mathbf{x}=(x_1,x_2,\ldots,x_n)\in \{0,1\}^n$, a deletion probability $p=k/n^{\alpha}$, where $k>1$ and $\alpha\in (0.5,1]$ are constants, and an integer $t$, the channel returns $t$ traces of~$\mathbf{x}$. Each trace of $\mathbf{x}$, denoted by $\mathbf{y}_i \in \{0,1\}^*$, $i=1,\ldots,t$, is obtained by sending $\mathbf{x}$ through a deletion channel with deletion probability $p$, i.e., the deletion channel deletes each bit $x_i$ of $\mathbf{x}$, $i=1,\ldots,n$, independently with probability $p$, and outputs a subsequence of $\mathbf{x}$ containing all bits of $\mathbf{x}$ that were not deleted in order. The $t$ traces $\mathbf{y}_1,\ldots,\mathbf{y}_t$, are independent and identically distributed as outputs of the deletion channel for input $\mathbf{x}$. The inputs of the reconstruction algorithm are the $t$ traces $\mathbf{y}_1,\ldots,\mathbf{y}_t$, the codeword length $n$, and the channel parameters $k$ and $\alpha$. Let $\hat{\mathbf{x}}\in \{0,1\}^n$ denote the output of the reconstruction algorithm. The probability of reconstruction error is defined by $P_e\triangleq \Pr(\hat{\mathbf{x}} \neq \mathbf{x})$, and the edit distance error is defined by the random variable $L_d \triangleq \lev(\mathbf{x},\hat{\mathbf{x}})$, where $\lev(.,.)$ denotes the Levenshtein distance~\cite{L66} between two sequences, and the randomness in $P_e$ and $L_d$ is over the deletion process. Let $\mathbf{a}$ and $\mathbf{b}$ be two binary sequences, we use the notation $\langle \mathbf{a},\mathbf{b} \rangle$ to refer to the concatenation of the two sequences $\mathbf{a}$ and $\mathbf{b}$. All logarithms in this paper are of base $2$ unless otherwise specified. 

Following standard notation, $f(n) = o(g(n))$ means $\lim_{n\to \infty} f(n)/g(n) = 0$; $f(n) = \omega(g(n))$ means $\lim_{n\to \infty} f(n)/g(n) =\infty$; $f(n) = \Omega(g(n))$ means $f$ is asymptotically bounded below by $c_1g(n)$ for some constant $c_1> 0$; $f(n) = O(g(n))$ means $f$ is asymptotically bounded above by $c_2g(n)$ for some constant $c_2> 0$; \mbox{$f(n)= \Theta(g(n))$} means $f(n)/g(n)$ asymptotically lies in an interval $[c_1, c_2]$ for some constants $c_1, c_2 > 0$, and $f(n)=\text{poly}(g(n))$ means $f(n)=\left( g(n)\right)^{O(1)}$.

\section{Code Construction}
 \label{sec:4:3}
 In this section, we present our code construction and explain the main ideas behind the techniques that we use. We also present our theoretical results in Section~\ref{code:full}.
 
\subsection{Overview}
\label{encodeco}

The high-level idea of our code construction is the following. We construct a codeword $\mathbf{x}\in \{0,1\}^n$, by concatenating $n/\ell$ strings each of length $\ell$, where $\ell \in \mathbb{Z}^+$ is a parameter of the code. For the sake of simplicity, we will assume that $\ell$ divides $n$ and drop this assumption later in Section~\ref{code:full}. Let $\mathbf{x}^m\in \{0,1\}^{\ell}$, $m\in \{1,\ldots,n/\ell\}$, be the concatenated strings which form the codeword $\mathbf{x}=\langle \mathbf{x}^1,\mathbf{x}^2,\ldots,\mathbf{x}^{n/\ell}\rangle\in \{0,1\}^n$. Henceforth, we use the term {\em blocks} to refer to $\mathbf{x}^1,\mathbf{x}^2,\ldots,\mathbf{x}^{n/\ell}$. We construct each block in $\mathbf{x}$ such that it satisfies the following two properties. First, the blocks satisfy a {\em run-length-limited} constraint, i.e., the length of each run of $0$'s or $1$'s is limited by a maximum value. Second, each block has a small number of fixed bits in the beginning and in the end of the block which we call {\em delimiter bits}. 

The goal of introducing the delimiter bits is to {\em detect} the exact number of deletions in each block and consequently recover the boundaries of the blocks at the decoder. Determining the block boundaries allows us to obtain the traces corresponding to each block. This enables us to subdivide the trace reconstruction problem into smaller subproblems. These subproblems are ``easier" in the sense that the expected number of deletions per block is smaller than the total number of deletions in $\mathbf{x}$, for the same number of traces. This increases the chance of reconstructing a higher number of segments of the codeword correctly, resulting in a low edit distance error as we show in Section~\ref{sec:simul}. Moreover, the delimiter bits can also prevent potential reconstruction errors from propagating to all parts of the codeword. To understand the intuition behind the subdivision, we present the following example.

Consider a naive trace reconstruction algorithm that attempts to reconstruct the sequence by finding a deletion-free copy of it among the traces. In the uncoded setting with a small deletion probability, we expect to observe a small number of deletions per trace. We also expect that many parts (i.e., blocks) of the trace will be deletion-free. Hence, if we use a code that consists of codewords of the form $\mathbf{x}=\langle \mathbf{x}^1,\mathbf{x}^2,\ldots,\mathbf{x}^{n/\ell}\rangle\in \{0,1\}^n$ that enables recovering the boundaries of each block of size $\ell$, then we have a higher chance of reconstructing the sequence by finding one deletion-free copy of each block among the traces. 

For instance, consider the setting where the decoder has $t=2$ traces of a binary sequence $\mathbf{x}$ of length $n$. Suppose that the first trace $\mathbf{y}_1$ is affected by a single deletion at position $i_1$ where $i_1<n/2$, and the second trace $\mathbf{y}_2$ is affected by a single deletion at position $i_2$ where $i_2>n/2$. Therefore, the decoder has
\begin{align*}
\mathbf{y}_1 &= (x_1,...,x_{i_1-1},x_{i_1+1},...,x_{n/2},............................,x_n), \\
\mathbf{y}_2 &= (x_1,............................,x_{n/2},...,x_{i_2-1},x_{i_2+1},...,x_n).
\end{align*}
If $\mathbf{x}$ is uncoded, the reconstruction is unsuccessful since there is no deletion-free copy of $\mathbf{x}$ in both traces. Now suppose that $\mathbf{x}$ is a codeword that belongs to a code that can detect up to 1 deletion in each of the $2$ blocks of size $\ell=n/2$. Then, $\mathbf{x}$ can be successfully reconstructed since knowing the boundaries of each block would allow us to obtain a deletion-free copy of the first block in the second trace, and a deletion-free copy of the second block in the first trace. Note that in general, we may have cases where we cannot find a deletion-free copy of every block. However, we expect that the subdivision will still allow us to find a deletion-free copy of most blocks, resulting in a lower edit distance error compared to the uncoded setting.

Motivated by the previous example, to reconstruct $\mathbf{x}$ at the decoder, we use the following approach. The decoder obtains $t$ independent traces of $\mathbf{x}$ resulting from $t$ independent deletion channels. To reconstruct $\mathbf{x}$, the decoder first uses the delimiter bits to recover the block boundaries of each trace, as we explain in Section~\ref{ch4:delimiter}. The decoder then uses the Bitwise Majority Alignment (BMA) algorithm which we explain in Section~\ref{ch4:bma} to reconstruct each block from its corresponding traces. To finalize the decoding, the recovered blocks are concatenated in order to reconstruct~$\mathbf{x}$. 

\subsection{Bitwise Majority Alignment}
\label{ch4:bma} 
The bitwise majority alignment (BMA) algorithm was first introduced in~\cite{Batu} as an algorithm that reconstructs an {\em uncoded} binary sequence $\mathbf{x}\in \{0,1\}^n$ from $t$ traces. In our approach, we first determine the boundaries of each block of size $\ell$ and then apply the BMA reconstruction algorithm at the level of each block. Hence, the input of the algorithm is a $t \times \ell$ binary matrix which consists of the $t$ traces corresponding to a certain block. Since the length of some of the received traces may be smaller than $\ell$ due to the deletions, the traces in the input of the algorithm are padded to length $\ell$ by adding a special character (other than $0$ or $1$). The main idea of the algorithm follows a majority voting approach, where each bit is reconstructed based on its value in the majority of the traces. Namely, for each trace, a pointer is initialized to the leftmost bit, and the value of the bit in the reconstructed sequence is decided based on the votes of the majority of the pointers. Then, the pointers corresponding to the traces that voted with the majority are incremented by one, while other pointers remain at the same bit position. Thus, the reconstruction process scans the bits of the traces from left to right, and the pointers may possibly be pointing to different bit positions at different states of the algorithm. In the end, the algorithm outputs a single reconstructed sequence of length $\ell$ bits. 

Let $q(j)$ be the pointer corresponding to trace $j$, $j=1,\ldots,t$. $q(j)\in \{1,\ldots,\ell\}$, where $q(j)=i$, $i=1,\ldots,\ell$, means that the pointer corresponding to trace $j$ is pointing to the bit at position $i$. The detailed algorithm is given in Algorithm~\ref{algo_disjdecomp}.

\IncMargin{1em}
\begin{algorithm}[h]
\SetKwInOut{Input}{input}\SetKwInOut{Output}{output}
\Input{A binary matrix $T$ of size $t\times \ell$}
\Output{A binary sequence $y$ of size $\ell$}
\BlankLine
Let $q(j) \leftarrow 1$ for all $j=1,\ldots,t$\;
\For{$i\leftarrow 1$ \KwTo $\ell$}{
	Let $b$ be the majority over all $j$ for $T(j,q(j))$\;
	$y(i)\leftarrow b$\;
	\For{$j\leftarrow 1$ \KwTo $t$}{
		\If{$q(j)==b$}{
			$q(j)\leftarrow q(j)+1$\; 
		}
	} 
}
\caption{Bitwise Majority Alignment}\label{algo_disjdecomp}
\end{algorithm}

The authors in~\cite{Batu} showed that for a deletion probability that satisfies $p=\mathcal{O}(1/\sqrt{\ell})$, the BMA algorithm (Algorithm~\ref{algo_disjdecomp}) can reconstruct an arbitrary (uncoded) binary sequence of size $\ell$, with high probability\footnote{The probability of reconstruction error vanishes asymptotically in $\ell$, where the probability is taken over the randomness of the deletion process.}, from $t=O(\ell \log \ell)$ traces. The BMA algorithm has a poor performance for some sequences that have {\em long} runs of $0$'s or $1$'s~\cite{Batu}. For example, consider a sequence that starts with a long run of $0$'s. Due to the deletions, the number of $0$'s observed for this run differs from one trace to another. While scanning from left to right, at some point before the end of this run, the majority of the traces could vote for a~$1$. This will lead to splitting the run resulting in an erroneous reconstruction. The successful reconstruction of such sequences with long runs requires a number of traces that is significantly larger than that of {\em run-length-limited} sequences. In fact, an intermediary result in~\cite{Batu} showed that for {\em run-length-limited} sequences where the length of any run is limited to a maximum value of $\sqrt{\ell}$, the number of traces required by the BMA algorithm is $t=\mathcal{O}(1)$. However, in the context of uncoded worst-case trace reconstruction considered in~\cite{Batu}, the BMA algorithm would still require $t=O(\ell \log \ell)$ traces so that the probability of reconstruction error (taken over the randomness of the deletion process) vanishes asymptotically for {\em any} binary sequence. In this paper, one of our approaches is to leverage coded trace reconstruction to decrease the required number of traces by restricting our code to codewords of the form $\mathbf{x}=\langle \mathbf{x}^1,\mathbf{x}^2,\ldots,\mathbf{x}^{n/\ell}\rangle\in \{0,1\}^n$, where the length of a run in any block of size $\ell$ is upper bounded by $\sqrt{\ell}$.

\begin{remark}
\label{remBMA}
A straightforward coded version of the BMA algorithm can be obtained by restricting the input codewords $\mathbf{x}\in \{0,1\}^n$  to sequences that do not contain a run that is greater than $\sqrt{n}$. Note that this simple variation of the BMA reconstruction algorithm has not been discussed in the literature, but based on an intermediary result in~\cite{Batu}, one can easily see that this coded version of the BMA algorithm would require a constant number of traces. However, our results in Section~\ref{sec:simul} show that there is a significant difference in terms of the edit distance error between the code that we propose in this paper and this simple coded version of the BMA algorithm. This highlights the importance of the step of introducing the delimiter bits to subdivide the problem into smaller subproblems which we will explain in the next section.
\end{remark}

\subsection{Detecting Deletions in Concatenated Strings}
\label{ch4:delimiter}
As previously mentioned, our code consists of codewords of the form $\mathbf{x}=\langle \mathbf{x}^1,\mathbf{x}^2,\ldots,\mathbf{x}^{n/\ell}\rangle\in \{0,1\}^n$ where each block $\mathbf{x}^m$, $m\in \{1,\ldots,n/\ell\}$, has some fixed delimiter bits. The goal of introducing the delimiter bits is to detect the exact number of deletions in each block, and hence enable the recovery of the blocks in each trace after the sequence $\mathbf{x}$ passes through the deletion channels. Designing codes that detect the number of deletions in concatenated strings is a problem of independent interest which was studied in~\cite{detecting}. The code construction presented in~\cite{detecting} for detecting up to $\delta$ deletions in each block $\mathbf{x}^m$, $m\in \{1,\ldots,n/\ell\}$, in $\mathbf{x}=\langle \mathbf{x}^1,\ldots,\mathbf{x}^{n/\ell}\rangle$ is given below.

\begin{construction}[Code detecting up to $\delta$ deletions~\cite{detecting}]
\label{cons1}
For all $\delta,\ell,n\in \mathbb{Z}^+$, with \mbox{$2\delta< \ell \leq n/2$}, we define the following%
\begin{align*}
\mathcal{A}_{\delta}^0(\ell) &\triangleq \big\{\mathbf{x}\in \{0,1\}^{\ell}~\big|~\mathbf{x}_{[1,\delta+1]}=\mathbf{0}^{\delta+1}\big\},\\
\mathcal{A}_{\delta}^1(\ell) &\triangleq \big\{\mathbf{x}\in \{0,1\}^{\ell} ~\big|~ \mathbf{x}_{[\ell-\delta+1,\ell]}=\mathbf{1}^{\delta}\big\}.
\end{align*}
The code $\mathcal{D}_{\delta}(\ell,n)\subseteq \mathbb{F}_2^n$ is defined as the set
\begin{equation*}
\left\{
  \langle \mathbf{x}^1,\ldots,\mathbf{x}^{n/\ell}\rangle \;\middle|\;
  \begin{aligned}
  & \mathbf{x}^1\in \mathcal{A}_{\delta}^1(\ell),\\
  & \mathbf{x}^m \in \mathcal{A}_{\delta}^1(\ell) \cap \mathcal{A}_{\delta}^0(\ell), \forall m \in [2, \frac{n}{\ell}-1], \\
  & \mathbf{x}^{n/\ell} \in \mathcal{A}_{\delta}^0(\ell).
  \end{aligned}
\right\}.
\end{equation*}
\end{construction}

The main result in~\cite{detecting} is summarized below. The authors in~\cite{detecting} also proved that the redundancy of the code $\mathcal{D}_{\delta}(\ell,n)$ given by Construction~\ref{cons1} is asymptotically optimal in $\delta$ among all codes that can detect up to $\delta$ deletions per block.

\begin{theorem}[\hspace{-0.008cm}\cite{detecting}]
\label{thm:d}
For $\delta,\ell,n\in \mathbb{Z}^+$, with $2\delta< \ell \leq n/2$, let $$\mathbf{x}=\langle \mathbf{x}^1,\mathbf{x}^2,\ldots,\mathbf{x}^{n/\ell}\rangle \in \mathcal{D}_{\delta}(\ell,n).$$ Suppose that $\mathbf{x}$ is affected by at most $\delta$ deletions in each of its blocks $\mathbf{x}^1,\ldots,\mathbf{x}^{n/\ell}$. The code $\mathcal{D}_{\delta}(\ell,n)$ given in Construction~\ref{cons1} detects up to $\delta$ deletions per block. This code is encodable and block-by-block decodable in linear time $O(n)$, and its redundancy is $(2\delta+1)(n/\ell-1)$ bits.
\end{theorem}
Next, we explain the idea behind the code $\mathcal{D}_{\delta}(\ell,n)$ by going through the decoding process for $\delta=1$.  Consider a codeword $\mathbf{x}\in \mathcal{D}_1$ that if affected by deletions resulting in a string $\mathbf{y}=(y_1,y_2,\ldots)$. To understand why $\mathcal{D}_{1}(\ell,n)$ can detect up to one deletion per block, we go over the possible values of the bit $y_{\ell}$. Based on the value of $y_{\ell}$, a decision rule can be formed which allows detecting the exact number of deletions in the first block (no deletion or $1$ deletion). As a result, the code enables determining the boundary of the first block and then the same decision rule can be applied  to detect the deletions in the remaining blocks by operating on a block-by-block basis. The possible values of   $y_{\ell}$ and the corresponding deletion scenarios are shown below.
\begin{enumerate}[leftmargin=*]
\item {\bf No deletion in first block:} By construction, we always have $y_{\ell}=1$ since the bits in the first block will not experience any shift in position.
\item {\bf One deletion in first block:} Notice that any single deletion in the first block will shift the first bit in the second block by one position to the left. By construction, the first two bits in the second block are $00$. Therefore, for up to one deletion in the second block, the value of the bit that will shift to position $\ell$ is always a $0$, i.e., $y_{\ell}=0$ for any single deletion in the first block given that at most one deletion occurs in the second block.
\end{enumerate}
Therefore, based on the value of $y_{\ell}$, the code can detect the number of deletions in each block with zero-error, given the assumption that each block is affected by at most one deletion. In the general case where $\delta>1$, the code can detect the exact number of deletions in each block given that at most $\delta$ deletions occur in each block. This is done by observing the values of the bits $y_{\ell-\delta},\ldots, y_{\ell}$. We omit the details of the general decoding algorithm and refer interested readers to~\cite{detecting}.

\begin{figure*}[!htb]
\minipage{0.333\textwidth}
  \includegraphics[width=\linewidth]{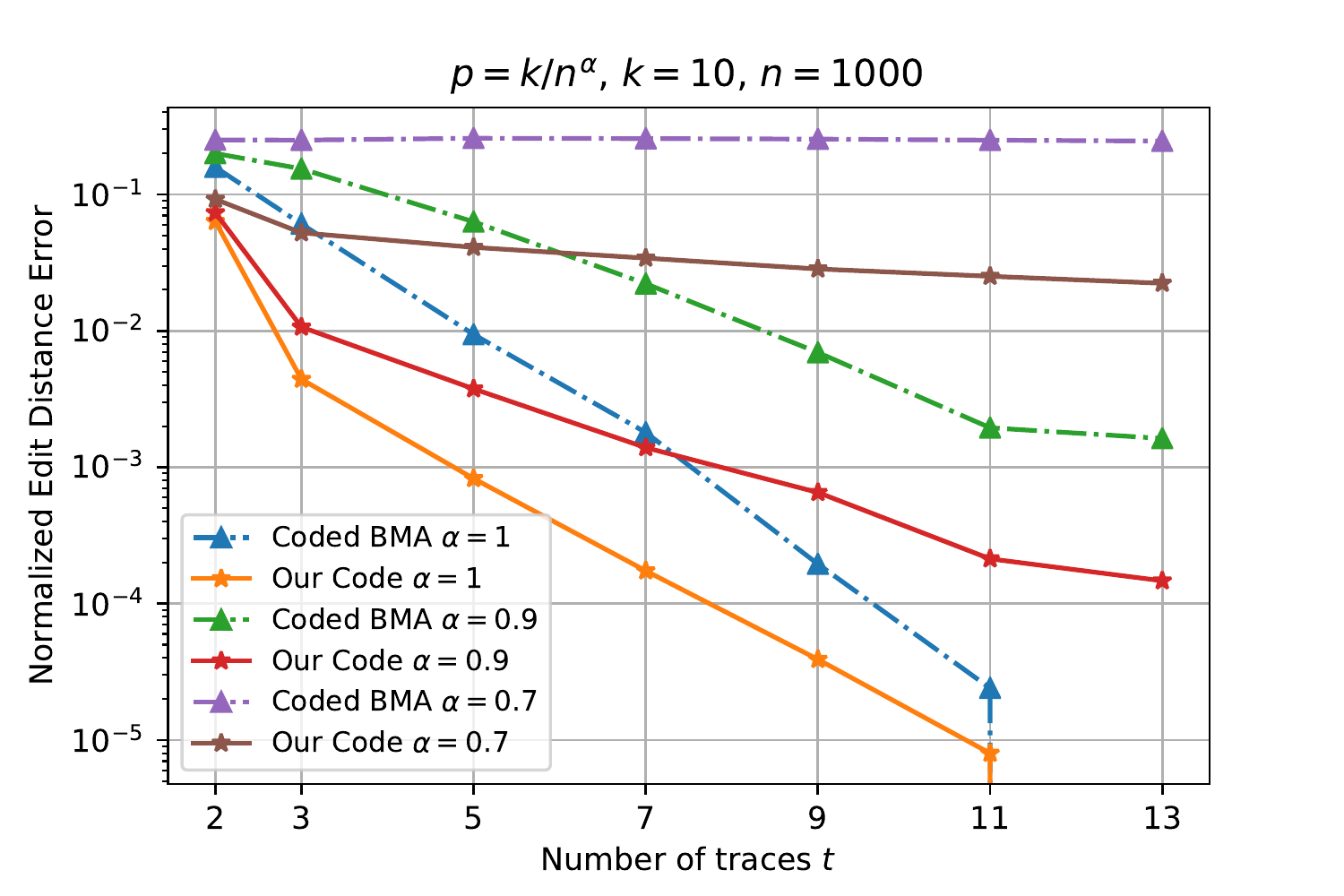}
  \caption{$p=k/n^{\alpha}$, $k=10$ and $n=1000$}\label{fig:1}
\endminipage\hfill
\minipage{0.333\textwidth}
  \includegraphics[width=\linewidth]{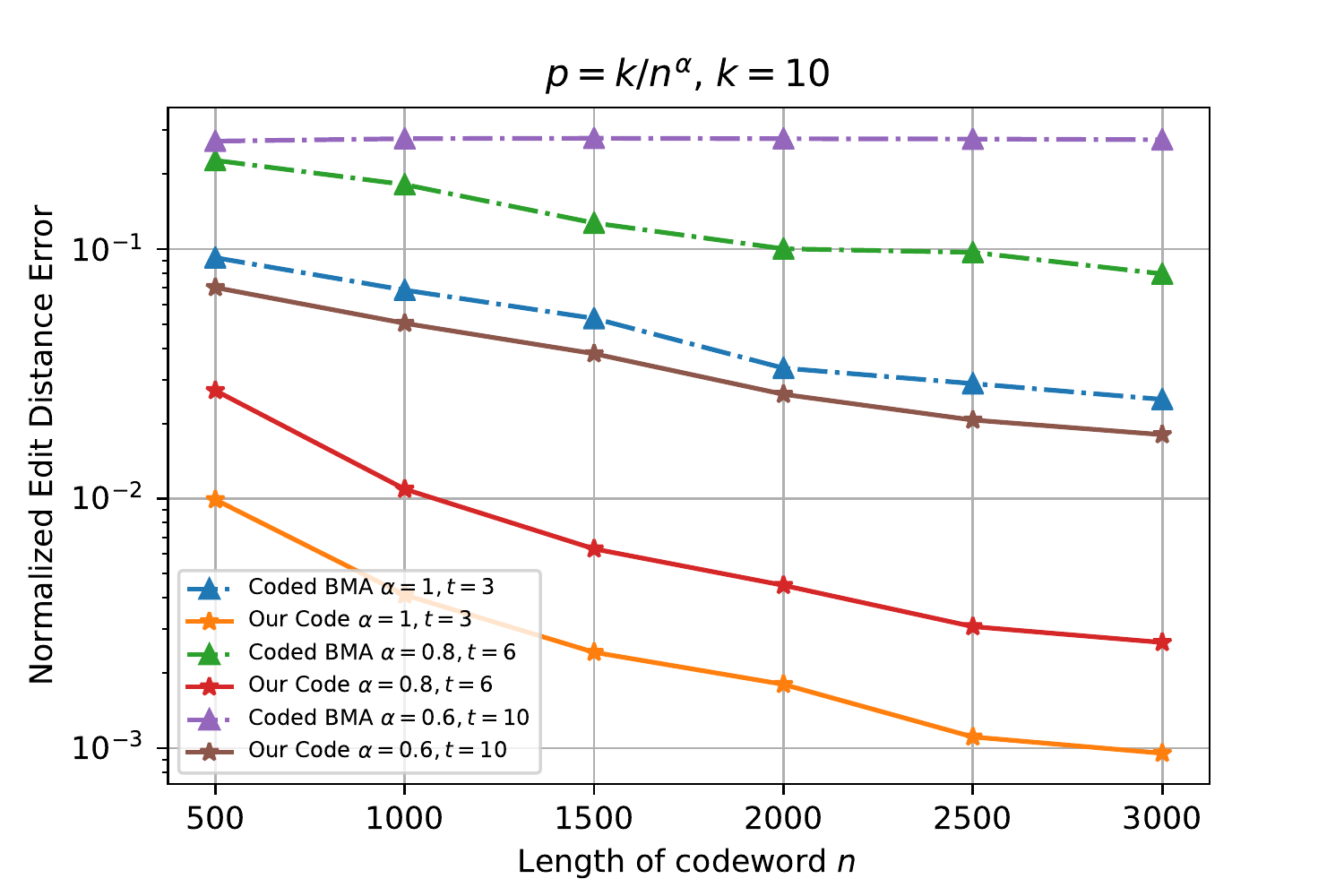}
  \caption{$p=k/n^{\alpha}$ and $k=10$}\label{fig:2}
\endminipage\hfill
\minipage{0.333\textwidth}%
  \includegraphics[width=\linewidth]{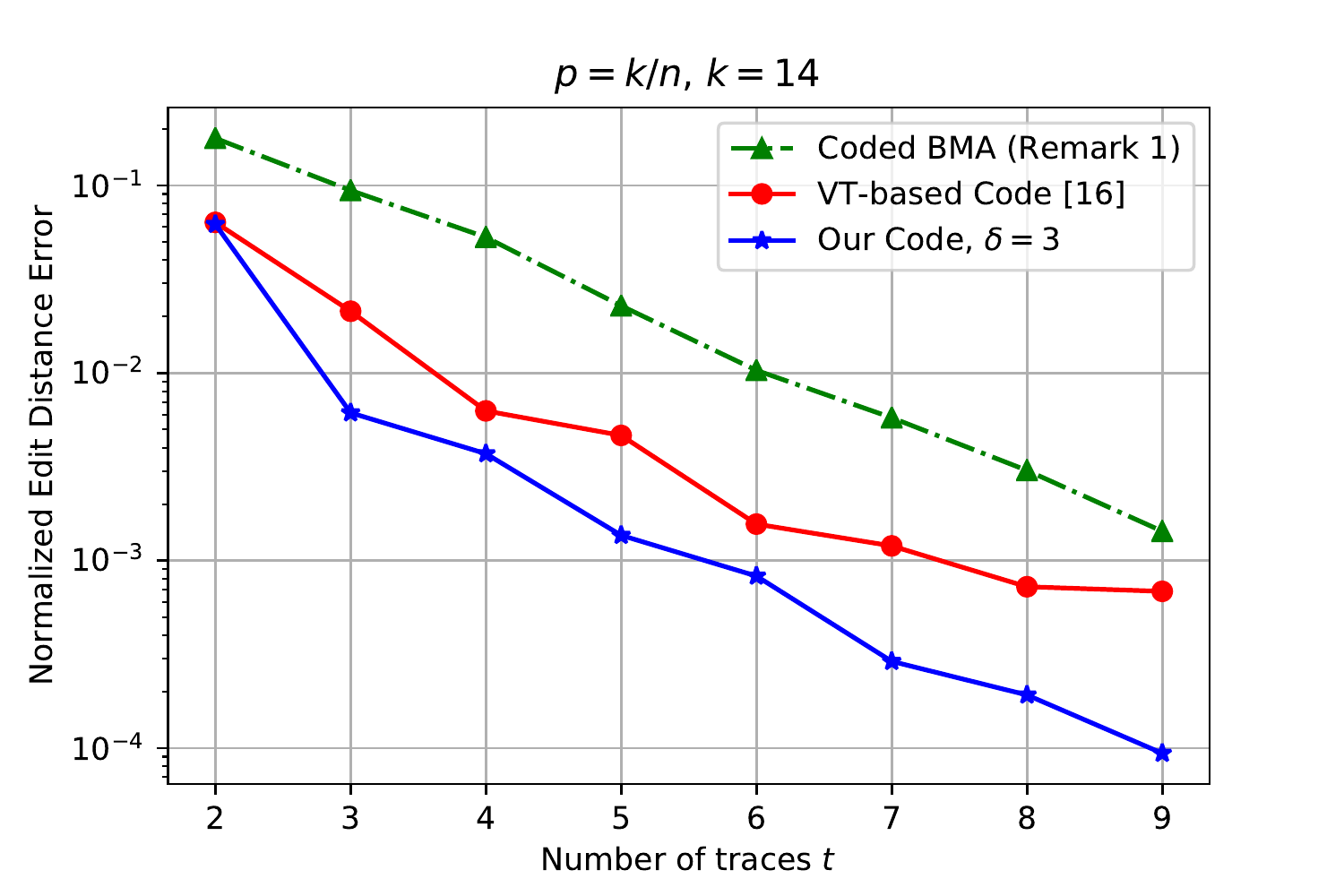}
  \caption{$p=k/n$, $k=14$, $n=994$}\label{fig:3}
\endminipage
\end{figure*}

\subsection{Theoretical Results}
\label{code:full}

Let $\mathcal{L}^{n}(f(n))\subseteq \{0,1\}^{n}$ be the set of {\em run-length-limited} binary sequences of length $n$ such that the length of any run of $0$'s or $1$'s in a sequence $\mathbf{x}\in \mathcal{L}^{n}(f(n))$ is at most $f(n)$.

\begin{construction}[Code for Trace Reconstruction]
\label{cons:full}
Consider a given deletion probability \mbox{$p=k/n^{\alpha}<1$}, where $k>1$ and \mbox{$\alpha\in(0.5,1]$} are constants. The code consists of codewords of the form $\mathbf{x}=\langle \mathbf{x}^1,\mathbf{x}^2,\ldots,\mathbf{x}^{\lceil n/\ell \rceil}\rangle$, where the length of each block is set to $\ell=\lfloor 1/p \rfloor$. Let $\delta \in \mathbb{Z}^+$, with $\delta>1$, be a code parameter that represents a strict upper bound on the number of deletions that can be detected in each block $\mathbf{x}^m$, $m\in \{1,\ldots, \lceil n/\ell \rceil\}$, i.e., the code can detect up to $\delta-1$ deletions per block. For $\ell=\lfloor 1/p \rfloor>\delta^2$ we define the code
$$\mathcal{C}_{\delta}(n)\triangleq \left \{ \mathbf{x}\in \{0,1\}^n~|~\mathbf{x}\in \mathcal{D}_{\delta-1}(\lfloor 1/p \rfloor,n) \cap \mathcal{L}^{n}(\sqrt{\lfloor 1/p \rfloor}) \right \}$$
\end{construction}

The steps of the reconstruction algorithm for a sequence $\mathbf{x}\in\mathcal{C}_{\delta}(n)$ are as follows.
\begin{enumerate}
\item Recover the block boundaries in all $t$ traces based on the decoding algorithm of $\mathcal{D}_{\delta}(\lfloor 1/p \rfloor,n)$.
\item Apply the BMA algorithm (Algorithm~\ref{algo_disjdecomp}) to reconstruct each block.
\item Concatenate the reconstructed blocks to obtain $\mathbf{x}$.
\end{enumerate}

\begin{theorem}
\label{thm:main1}
For any deletion probability \mbox{$p=k/n^{\alpha}$} with $0<p<0.5$, where $k>1$ and \mbox{$\alpha\in(0.5,1]$} are constants, the code $\mathcal{C}_{\delta}(n)\subseteq \{0,1\}^n$ (Construction~\ref{cons:full}) can be reconstructed from $t=O(1)$ traces in linear time $O(n)$. The redundancy of the code $r_{\mathcal{C}}(\delta)$ satisfies $$(kn^{1-\alpha}-1)(2\delta-1)+o(1)\leq r_{\mathcal{C}}(\delta) < 2kn^{1-\alpha}(2\delta-1)+o(1).$$ Let $p(n)=O( \text{{\em poly}} (n))$, for $\delta=\lceil \delta^*(p(n)) \rceil=o(\log n)$ with $$\delta^*(p(n))=\frac{2\ln \left( \sqrt{e}n^{1-\alpha}p(n) \right)}{W\left(2e \ln \left( \sqrt{e} n^{1-\alpha}p(n)  \right) \right)},$$ where $W(.)$ is the Lambert $W$ function, the probability of reconstruction error for {\em any} codeword $\mathbf{x}\in \mathcal{C}_{\delta}(n)$ satisfies $$P_e < O\left(n^{1-2\alpha} + \frac{1}{p(n)}\right).$$
 \end{theorem}
 \begin{corollary}
 \label{cor1}
 For any deletion probability \mbox{$p=k/n^{\alpha}$} with \mbox{$0<p<0.5$}, where $k> 1$ and \mbox{$\alpha\in(0.5,1)$} are constants, the code defined in Construction~\ref{cons:full} can be reconstructed from $O(1)$ traces in linear time $O(n)$, with redundancy $$r_{\mathcal{C}}
=\Theta \left(n^{1-\alpha} \frac{\log n}{\log \log n} \right),$$ and a probability of reconstruction error that satisfies $$P_e< O(n^{1-2\alpha}),$$ for any codeword.
 \end{corollary}
 \begin{corollary}
 \label{cor2}
For any deletion probability \mbox{$p=k/n$} with \mbox{$0<p<0.5$}, where $k> 1$ is a constant, the code defined in Construction~\ref{cons:full} can be reconstructed from $O(1)$ traces in linear time $O(n)$, with a probability of reconstruction error that vanishes asymptotically in $n$ for any redundancy $r_{\mathcal{C}}=\omega(1)$.
 \end{corollary}
 Theorem~\ref{thm:main1} shows that the code defined in Construction~\ref{cons:full} can be efficiently reconstructed from a constant number of traces, and also expresses the redundancy and the probability of reconstruction error in terms of the code parameters and shows a relationship (trade-off) between these two code properties. Corollaries~\ref{cor1} and \ref{cor2} follow from Theorem~\ref{thm:main1}. Corollary~\ref{cor1} shows that for $\alpha\in(0.5,1)$ the redundancy is $$r_{\mathcal{C}}=\Theta \left(n^{1-\alpha}\frac{\log n}{\log \log n}\right)=o(n^{1-\alpha}\log n),$$ and the probability of reconstruction error vanishes asymptotically in $n$ as a polynomial of degree $1-2\alpha$. Note that the aforementioned redundancy grows strictly slower than the optimal redundancy of a deletion correcting code (with $t=1$ trace); moreover, the complexity of reconstruction is linear in~$n$. Corollary~\ref{cor2} shows that for $\alpha=1$, a vanishing probability of reconstruction error can be achieved with  a constant number of traces and any $\omega(1)$ redundancy.

\section{Simulation Results and Comparison}
\label{sec:simul}
In this section, we evaluate the numerical performance of our code $C_{\delta}(n)$ defined in Construction~\ref{cons:full} in terms of the edit distance error $L_d=\lev (\hat{\mathbf{x}},\mathbf{x})$ \cite{Diggavi, Eitan, Approximate}. We provide simulation results where we compare the normalized edit distance error $\lev (\hat{\mathbf{x}},\mathbf{x})/n$ of our code to: \begin{enumerate*}[label={\textit{(\roman*)}}] \item The coded version of the BMA algorithm (Remark~\ref{remBMA}) for multiple value of $t, n$ and \mbox{$\alpha\in(0.5,1]$}, and \item The code in~\cite{AbroshanTrace} for $\alpha=1$ and multiple values of $t$. \end{enumerate*}

The simulation results are shown in Fig.~\ref{fig:1}, \ref{fig:2}, and \ref{fig:3}. The edit distance error is averaged over $10^3$ runs of simulations, and then normalized by the blocklength $n$. In each run, a codeword of length $n$ is chosen uniformly at random, and the random deletion process is applied to obtain the $t$ traces. The parameter $\delta$ of the code $C_{\delta}(n)$ is set to $\delta=3$ for all simulations. The results, in general, show that our code has a much lower edit distance error compared to the simple coded version of the BMA~(Remark~\ref{remBMA}). This confirms our intuition about the importance of introducing the delimiter bits and subdividing the problem of reconstructing $\mathbf{x}$ into smaller subproblems. More specifically, the subproblems are easier since they deal with a lower number of deletions; and the delimiter bits prevent potential reconstruction errors from propagating to the entire codeword resulting in a lower edit distance error. For instance, we observe in Fig.~\ref{fig:2} that for $n=3000$ the normalized edit distance error is $\approx 10^{-3}$ for our code and $\approx 2.5 \times 10^{-2}$ for the coded BMA. Note that this improvement comes at the expense of a lower code rate; however, different trade-offs can be achieved by varying the parameters $\delta$ and $\ell$. 

In Fig.~\ref{fig:3}, we also compare the performance our code to the VT-based code in~\cite{AbroshanTrace} in the regime where the average number of deletions per segment is a constant (i.e., $\alpha=1$). The code in \cite{AbroshanTrace} is constructed by concatenating $n/\ell=17$ VT codewords each of size $\ell=59$, resulting in a code with rate~$\approx 0.9$. The parameters of our code are set to $\delta=3$, $n=994$, and $\ell=71$, resulting in a code with rate $\approx 0.934$. The results show that our code has a lower edit distance error compared to the code in~\cite{AbroshanTrace}, for a higher code rate. 

\section{Proofs}
\label{proof}
\subsection{Proof of Theorem~\ref{thm:main1}}
First, we derive an expression of the redundancy of the code $\mathcal{C}_{\delta}(n)$ defined in Construction~\ref{cons:full}, denoted by $r_{\mathcal{C}}(\delta)$, in terms of the code parameter~$\delta$, where $\delta=o(\log n)$. It follows from Theorem~\ref{thm:d} that the redundancy corresponding to $\mathcal{D}_{\delta-1}(\lfloor 1/p \rfloor,n)$ (Construction~\ref{cons1}), denoted by $r_{\mathcal{D}}(\delta)$, is 
\begin{equation}
\label{eqq1}
r_{\mathcal{D}}(\delta)=(2\delta-1)\left(\left \lceil \frac{n}{\ell} \right \rceil-1\right), 
\end{equation}
where based on Construction~\ref{cons:full} the size of each block is $\ell=\lfloor 1/p \rfloor$, and the number of blocks is $\lceil n/\ell \rceil$. The following claim gives lower and upper bounds on the number of blocks by applying standard bounds on the floor and the ceiling functions. 
\begin{claim}
\label{claim1}
For $n>1$ and $p<1/2$, we have
\begin{align}
kn^{1-\alpha}-1 &\leq \left \lceil \frac{n}{\ell} \right \rceil -1 < 2kn^{1-\alpha}, \label{eqq2} \\
kn^{1-\alpha} &\leq \left \lceil \frac{n}{\ell} \right \rceil < (2k+1)n^{1-\alpha}. \label{eqq3}
\end{align}
\end{claim}
We assume that the claim is true in what follows and give its proof in Section~\ref{sec:proofclaim}. By applying the bound in~\eqref{eqq2} to the expression of $r_{\mathcal{D}}(\delta)$ in~\eqref{eqq1} we get
\begin{equation}
\label{eq:bb}
(kn^{1-\alpha}-1)(2\delta-1)\leq r_{\mathcal{D}}(\delta) < 2kn^{1-\alpha}(2\delta-1).
\end{equation}
Next, we show that {\em run-length-limited} constraint in Construction~\ref{cons:full} introduces an additional redundancy that vanishes asymptotically in $n$, i.e., $r_{\mathcal{C}}(\delta)=r_{\mathcal{D}}(\delta)+o(1)$. 

Let $X\in \{0,1\}^n$ be a random variable that represents a binary sequence of length $n$ that is chosen uniformly at random. Let $L(X): \{0,1\}^n\to \{1,2,\ldots,n\}$ denote the length of the longest run in $X$. Recall that $\mathcal{L}^{n}(\sqrt{\ell})$ denotes the set of {\em run-length-limited} binary sequences of length $n$ such that the length of any run of $0$'s or $1$'s in a sequence $\mathbf{x}\in \mathcal{L}^{n}(\sqrt{\ell})$ is at most $\sqrt{\ell}$.
We define the following probabilities
\begin{align}
P_1&\triangleq P(X\in \mathcal{D}_{\delta-1}(\ell,n),X \notin \mathcal{L}^n(\sqrt{\ell})), \\
P_2&\triangleq P(X\in \mathcal{D}_{\delta-1}(\ell,n),X \in \mathcal{L}^n(\sqrt{\ell})),
\end{align}
where $\ell=\lfloor 1/p \rfloor = \lfloor n^{\alpha}/k \rfloor$. Note that 
\begin{equation}
P_2=P(X\in \mathcal{C}_{\delta}(n))=P(X\in \mathcal{D}_{\delta-1}(\ell,n))-P_1.
\end{equation}
As a first step, we are interested in computing an upper bound on $P_1$. For the sake of brevity, we will write $\mathcal{D}_{\delta-1}$ instead of $\mathcal{D}_{\delta-1}(\ell,n)$ in what follows. 
\begin{align}
P_1 &= P(X\in \mathcal{D}_{\delta-1},L(X)>\sqrt{\ell}) \\
&= P(L(X)>\sqrt{\ell}~\big|X\in \mathcal{D}_{\delta-1}) P(X\in \mathcal{D}_{\delta-1}) \\
&\leq n\cdot \frac{2}{2^{\sqrt{\ell}-\delta-1}}\cdot \frac{1}{2^{r_{\mathcal{D}}(\delta)}} \\
&=2^{-r_{\mathcal{D}}(\delta)}\cdot \frac{n}{2^{\sqrt{\ell}-\delta-2}}.
\end{align}
Therefore, 
\begin{align}
P_2 &= P(X\in \mathcal{D}_{\delta-1})-P_1 \\
&\geq 2^{-r_{\mathcal{D}}(\delta)} - 2^{-r_{\mathcal{D}}(\delta)}\cdot \frac{n}{2^{\sqrt{\ell}-\delta-2}} \\
&= 2^{-r_{\mathcal{D}}(\delta)} \left[1- \frac{n}{2^{\sqrt{\ell}-\delta-2}}  \right] \\
&= 2^{-r_{\mathcal{D}(\delta)}} \left[\frac{2^{\sqrt{\ell}-\delta-2}-n}{2^{\sqrt{\ell}-\delta-2}}  \right].
\end{align}
Hence,
\begin{equation}
|C_{\delta}(n)|=2^n \cdot P_2\geq 2^{n-r_{\mathcal{D}(\delta)}} \left[\frac{2^{\sqrt{\ell}-\delta-2}-n}{2^{\sqrt{\ell}-\delta-2}}  \right].
\end{equation}
Thus, we obtain the following upper bound on the redundancy
\begin{equation}
r_{\mathcal{C}}(\delta)\leq r_{\mathcal{D}}(\delta) + \log_2 \left[\frac{2^{\sqrt{\ell}-\delta-2}}{2^{\sqrt{\ell}-\delta-2}-n}  \right].
\end{equation}
Since $\ell=\lfloor 1/p \rfloor = \lfloor n^{\alpha}/k \rfloor$ where $k\geq 1$ and \mbox{$\alpha\in(0.5,1]$} are constants, then we have $\ell=\Theta(n^{\alpha})$. Furthermore, since $\delta=o(\log n)$, it follows that
\begin{equation}
\lim_{n\to +\infty} \log \left[\frac{2^{\sqrt{\ell}-\delta-2}}{2^{\sqrt{\ell}-\delta-2}-n}  \right] = \log(1)=0.
\end{equation}
Therefore,
\begin{equation}
r_{\mathcal{C}}(\delta)\leq r_{\mathcal{D}}(\delta) + o(1).
\end{equation}
Also, $r_{\mathcal{C}}(\delta)\geq r_{\mathcal{D}}(\delta)$ is a trivial lower bound. Therefore, we have
\begin{equation}
\label{eq:rr}
r_{\mathcal{C}}(\delta)= r_{\mathcal{D}}(\delta) + o(1).
\end{equation}
By combining the results in~\eqref{eq:rr} and \eqref{eq:bb} we conclude the proof for the redundancy of the code given in Theorem~\ref{thm:main1}.

Next, we derive an upper bound on the probability of error, i.e., the probability that the reconstruction is unsuccessful. We define the following events. Let $D$ denote the event where the process of detecting the deletions and determining the boundaries of the blocks is successful for all the blocks and all the $t$ traces. Let $B$ denote the event where the BMA algorithm (Algorithm~\ref{algo_disjdecomp}) is successful in reconstructing $\mathbf{x}=\langle \mathbf{x}^1,\mathbf{x}^2,\ldots,\mathbf{x}^{\lceil n/\ell \rceil}\rangle$, i.e., successful in reconstructing all the blocks $\mathbf{x}^1,\mathbf{x}^2,\ldots,\mathbf{x}^{\lceil n/\ell \rceil}$. Let $P_e$ denote the probability of error corresponding to Construction~\ref{cons:full}. For the sake of deriving an upper bound, we assume that the reconstruction is unsuccessful if the event $D$ is not realized. Hence, we can write
\begin{align}
P_e &\leq P(D^c) + P(B^c,D) \\
&= P(D^c)+P(B^c|D)P(D) \\
&\leq P(D^c)+P(B^c|D). \label{eq:perror}
\end{align}
Recall that in Construction~\ref{cons:full} we set $\ell=\lfloor 1/p \rfloor=\lfloor n^{\alpha}/k \rfloor$, and hence $p=\Theta(1/\ell)$. In the case where the process of determining the block boundaries is successful for all traces, it follows from~\cite{Batu} that the probability of error of the BMA algorithm for each block is\footnote{The constant in the upper bound on the probability $O(1/\ell)$ is not explicitly given in~\cite{Batu}.} $O(1/{\ell})$, for $p=\Theta(1/\ell)$, a constant number of traces, and an arbitrary sequence that does not have any run of length greater than $\sqrt{\ell}$. Therefore, by applying the union bound over the $\lceil n/\ell \rceil=\lceil n/\lfloor 1/p \rfloor \rceil$ blocks we obtain
\begin{align}
P(B^c|D) &\leq \left \lceil \frac{n}{\ell} \right \rceil O\left( \frac{1}{\ell} \right) \\
&< (2k+1)n^{1-\alpha} O\left( n^{-\alpha} \right) \\
&= O\left(n^{1-2\alpha} \right).  \label{eq:pbma}
\end{align}
Since $\alpha\in (0.5,1]$, it is easy to see from~\eqref{eq:pbma} that $P(B^c|D)$ vanishes asymptotically in $n$. Therefore, in the case where the block boundaries are recovered successfully at the decoder using the delimiter bits, the reconstruction is successful with high probability, where the probability is over the randomness of the deletion process. 

Next, we derive an upper bound on $P(D^c)$. Let \mbox{$Y_m, m\in \{1,\ldots,\lceil n/\ell \rceil\}$}, be the random variable that represents the exact number of deletions in block $m$ after $\mathbf{x}$ passes through the deletion channel with deletion probability~$p$. Note that \mbox{$Y_m \sim$ Binomial$(\ell,p)$}. It follows from the construction that the boundaries of a given block are recovered successfully if and only if $Y_m<\delta$, where $\delta=o(\log n)$ is a code parameter. Let $D_j, j\in \{1,\ldots,t\}$, denote the event where all the block boundaries of trace $j$ are determined successfully. Therefore, by applying the union bound over the number of blocks  $\lceil n/\ell \rceil$ blocks we obtain
\begin{align}
\label{eq:djc}
P(D_j^c)\leq \left \lceil \frac{n}{\ell} \right \rceil  P(Y_m \geq \delta).
\end{align}
Next, we upper bound $P(Y_m \geq \delta)$ by applying the following Chernoff bound on Binomial random variables.
\begin{proposition}[\hspace{-0.008cm}\cite{mitzen}]
\label{prop1}
Let $Y$ be a binomial random variable with $E[Y]=\mu$. For any $\delta>0$,
$$P(Y\geq (1+\delta)\mu)\leq \exp\left[-\mu\left((1+\delta)\ln(1+\delta)-\delta \right)\right].$$
\end{proposition}
\noindent Since $E[Y_m]=\ell p =\lfloor 1/p \rfloor p$, $p\leq 1/2$, and $\lfloor 1/p \rfloor \geq 1/p - 1$, then
\begin{equation}
E[Y_m]= \left \lfloor \frac{1}{p}\right \rfloor p \geq \left(\frac{1}{p}-1\right)p = 1-p \geq \frac{1}{2}.
\end{equation}
So by applying Proposition~\ref{prop1} to the Binomial random variable $Y_m$ we get
\begin{align}
P(Y_m\geq \delta) &\leq P(Y_m\geq \delta E[Y_m]) \\
&\leq e^{-E[Y_m]\left(\delta \ln \delta -\delta +1 \right)}\\
&\leq e^{-\frac{1}{2} \left(\delta \ln \delta - \delta +1 \right)}. \label{eq:25}
\end{align}
Therefore, by substituting~\eqref{eq:25} in \eqref{eq:djc} we get
\begin{align}
P(D_j^c) &\leq \left \lceil \frac{n}{\ell} \right \rceil  e^{-\frac{1}{2} \left(\delta \ln \delta - \delta +1 \right)} \\
&< (2k+1) n^{1-\alpha} e^{-\frac{1}{2} \left(\delta \ln \delta - \delta +1 \right)}, \label{eq:27}.
\end{align}
Let
\begin{equation}
\delta^*\left( p(n)\right) \triangleq \frac{2\ln \left( \sqrt{e}  n^{1-\alpha} p(n) \right)}{W\left(2e \ln \left( \sqrt{e} n^{1-\alpha} p(n)  \right) \right)},
\end{equation}
where $W(z)$ is the Lambert $W$ function defined as the solution of the equation $we^w=z$ with respect to the variable $w$. One can easily show that by substituting $\delta$ by $\delta^*(p(n))$ in~\eqref{eq:27} we get
\begin{equation}
\label{eqq:35}
P(D_j^c) < \frac{2k+1}{p(n)}.
\end{equation}
Since $\delta$ is an integer by construction, we set $\delta=\lceil \delta^*(p(n)) \rceil$, and the bound in~\eqref{eqq:35} still holds. Furthermore, since we consider a constant number of traces $t=O(1)$, then by applying the union bound over the $t$ traces we get
\begin{equation}
P(D^c) \leq \sum_{j=1}^{t} P(D_j^c) < t \cdot \frac{2k+1}{p(n)} = O\left(\frac{1}{p(n)}\right).
\label{eq:dc}
\end{equation}
Therefore, by substituting \eqref{eq:pbma} and \eqref{eq:dc} in \eqref{eq:perror} we obtain 
\begin{equation}
P_e < O\left(n^{1-2\alpha} + \frac{1}{p(n)}\right).
\end{equation}
Furthermore, the time complexity of the deletion detection algorithm is linear in $n$\cite{detecting}, and the complexity of the BMA algorithm is linear in the size of input matrix $t\times \ell$ since every trace is scanned once in the reconstruction process. Hence, the complexity of the reconstruction algorithm corresponding to Construction~\ref{cons:full} is linear in $n$, which concludes the proof of Theorem~\ref{thm:main1}.

\subsection{Proof of Corollary~\ref{cor1}}
The result in Corollary~\ref{cor1} follows from Theorem~\ref{thm:main1} for \mbox{$\alpha \in (0.5,1)$} by substituting $p(n)=n^{2\alpha-1}$ in the expression of $\delta^*$ and also in the bound on the probability of error $P_e$; in addition to applying standard bounds on the Lambert $W$ function. For $p(n)=n^{2\alpha-1}$, we have
\begin{equation}
\delta^*=\frac{2\ln \left( \sqrt{e}n^{\alpha} \right)}{W\left(2e \ln \left( \sqrt{e} n^{\alpha}  \right) \right)},
\end{equation}
\begin{equation}
\label{eq:tb}
P_e < O\left(n^{1-2\alpha} + \frac{1}{n^{2\alpha-1}}\right)= O\left( n^{1-2\alpha} \right),
\end{equation}
where the last equality in~\eqref{eq:tb} follows from the fact that the number of traces $t$ is a constant. Since $\alpha$ is a constant, we have $$2\ln \left( \sqrt{e}n^{\alpha} \right)=\ln e + 2\alpha \ln n= \Theta(\log n).$$ Furthermore, the following bound on the Lambert $W$ function can be found in~\cite{lambert}. For every $x\geq e$,
\begin{equation}
\ln x - \ln \ln x \leq W(x)  \leq \ln x -\frac{1}{2} \ln \ln x.
\end{equation}
It follows from the bound that $W(x)=\Theta(\ln x)$, so 
\begin{equation}
W\left(2e \ln \left( \sqrt{e} n^{\alpha}  \right) \right)= W(\Theta(\log n))=\Theta(\log \log n).
\end{equation} 
Therefore, we have $\delta^*=\Theta(\log n/\log \log n)$. Finally, from the bounds on $r_{\mathcal{C}}(\delta^*)$ in Theorem~\ref{thm:main1} we can deduce that \mbox{$r_{\mathcal{C}}=\Theta(n^{1-\alpha}\log n/ \log \log n)$}.

\subsection{Proof of Corollary~\ref{cor2}}
The result in Corollary~\ref{cor2} follows from Theorem~\ref{thm:main1} for the case where $\alpha=1$. For $\alpha=1$, we have
\begin{equation}
\delta^*=\frac{2\ln \left( \sqrt{e}p(n) \right)}{W\left(2e \ln \left( \sqrt{e} p(n)  \right) \right)},
\end{equation}
\begin{equation}
P_e < O\left(\frac{1}{n}+\frac{1}{p(n)}\right),
\end{equation}
and $r_{\mathcal{C}}(\delta^*)=\Theta(\delta^*)$. Hence, it is easy to see that for \mbox{$p(n)=\omega(1)$}, the probability of error vanishes asymptotically in $n$, where $r_{\mathcal{C}}=\omega(1)$.

\subsection{Proof of Claim~\ref{claim1}}
\label{sec:proofclaim}
\noindent For $x\in \mathbb{R}$, we have
\begin{align}
x-1 &<\lfloor x \rfloor \leq x \\
x &\leq \lceil x \rceil < x+1.
\end{align}
Therefore,
\begin{align}
\left \lceil \frac{n}{\ell} \right \rceil-1 &= \left \lceil \frac{n}{\lfloor 1/p \rfloor} \right \rceil -1 \\
 &< \left \lceil \frac{n}{\frac{1}{p}-1} \right \rceil - 1 \\
 &< \frac{n}{\frac{1}{p}-1} \\
 &= \frac{np}{1-p} \\
 &< 2np \label{eq59}\\ 
 &= 2kn^{1-\alpha},
\end{align}
where~\eqref{eq59} follows from the fact that $p<1/2$. Hence,
\begin{equation}
\left \lceil \frac{n}{\ell} \right \rceil <2k^{1-\alpha}+1\leq (2k+1)n^{1-\alpha},
\end{equation}
where the last inequality holds for $n\geq 1$ and is achieved with equality for $n=1$. Furthermore,
\begin{equation}
\left \lceil \frac{n}{\ell} \right \rceil = \left \lceil \frac{n}{\lfloor 1/p \rfloor} \right \rceil \geq \lceil np \rceil \geq np = kn^{1-\alpha}. 
\end{equation}

\bibliographystyle{ieeetr}

\bibliography{Refs4}
\end{document}